\documentclass[12pt,a4paper]{article}
\usepackage{amsfonts,latexsym}
\usepackage{amsmath,amssymb}
\usepackage{graphicx,color}
\usepackage{multirow}

\renewcommand{\thefootnote}{\fnsymbol{footnote}}

\begin{document}

\vspace{12mm}

\begin{center}
{{{\Large {\bf Scale-invariant tensor spectrum  from conformal  gravity}}}}\\[10mm]

{Yun Soo Myung\footnote{e-mail address: ysmyung@inje.ac.kr} and Taeyoon Moon\footnote{e-mail address: tymoon@inje.ac.kr}}\\[8mm]

{Institute of Basic Sciences and Department  of Computer Simulation, Inje University Gimhae 621-749, Korea\\[0pt]}

\end{center}
\vspace{2mm}

\begin{abstract}
We study cosmological tensor perturbations generated during de
Sitter inflation in the conformal gravity with mass parameter
$m^2=2H^2$. It turns out that tensor power spectrum is
scale-invariant.

\end{abstract}
\vspace{5mm}



\vspace{1.5cm}

\hspace{11.5cm}{Typeset Using \LaTeX}
\newpage
\renewcommand{\thefootnote}{\arabic{footnote}}
\setcounter{footnote}{0}


\section{Introduction}

It is well-known that the Einstein-Weyl gravity has seven
propagating gravitational modes, even the massive spin-2 state is a
ghost and it violates unitarity~\cite{Stelle:1976gc}.   The
conformal gravity  of
$\sqrt{-g}C^{\mu\nu\rho\sigma}C_{\mu\nu\rho\sigma}(=C^2)$ is
invariant under the Weyl transformation of $g_{\mu\nu}\to
\Omega^2(x)g_{\mu\nu}$ and thus, it has its own interests in gravity
and cosmology~\cite{Riegert:1984zz}. The Weyl invariance removes one
of seven gravitational modes and it propagates only six
modes~\cite{Deser:2012qg,Deser:2013bs,Hassan:2013pca}. In the flat
background, these are a healthy massless spin-2, a ghost massless
spin-2 and a massless spin-1.

On the other hand, the Einstein-Weyl gravity provides surely massive
scalar and vector propagations generated during de Sitter inflation
in addition to massive tensor ghost and massless
tensor~\cite{Clunan:2009er,Deruelle:2010kf,Deruelle:2012xv,Myung:2014jha}.
The authors have shown that in the limit of $m^2 \to 0$ (keeping the
conformal gravity only), the vector and tensor power spectra became
constant~\cite{Myung:2014cra}. Recently, the Harrison-Zel'dovich
scale-invariant spectrum was obtained from the Lee-Wick scalar model
in de Sitter spacetime~\cite{Myung:2014mla}. This model is a
fourth-order scalar theory whose mass parameter is given by
$M^2=2H^2$.

It is very important to look for  a scale-invariant tensor spectrum
which is the most robust and model-independent prediction of
inflation. We note that the tensor amplitude is a direct measure of
the expansion rate $H$ during inflation. This is in contrast to the
scalar amplitude which depends on both $H$ and $\varepsilon$ in the
slow-roll inflation. It is worth noting that the Einstein-gravity
provides a scale-invariant spectrum in the superhorizon region of
$k\ll aH(z\ll1)$ only. This is known to be a scale-invariant
superhorizon spectrum of tensor perturbation.

 In this Letter, we will find  a scale-invariant tensor power
spectrum generated during  de Sitter inflation from conformal
gravity with mass parameter $m^2=2H^2$.

\section{Einstein-Weyl gravity }

We start with  the Einstein-Weyl  gravity whose action is given by
\begin{equation} \label{ECSW}
S_{\rm EW}=\frac{1}{2\kappa}\int d^4x \sqrt{-g}\Big[ R-2\kappa
\Lambda-\frac{1}{2m^2}C^{\mu\nu\rho\sigma}C_{\mu\nu\rho\sigma}\Big],
\end{equation}
where the Weyl-squared term takes the form
\begin{eqnarray}
C^{\mu\nu\rho\sigma}C_{\mu\nu\rho\sigma}&=&2\Big(R^{\mu\nu}R_{\mu\nu}-\frac{1}{3}R^2\Big)+
(R^{\mu\nu\rho\sigma}R_{\mu\nu\rho\sigma}-4R^{\mu\nu}R_{\mu\nu}+R^2).
\end{eqnarray}
In the limit of $m^2\to 0$, one recovers the conformal gravity
action~\cite{Mannheim:2011is,Myung:2014cra}
\begin{equation} \label{CG}
S_{\rm CG}=-\frac{1}{4\kappa m^2}\int d^4x
\sqrt{-g}\Big[C^{\mu\nu\rho\sigma}C_{\mu\nu\rho\sigma}\Big].
\end{equation}
 Here we have
$\kappa=8\pi G=1/M^2_{\rm P}$, $M_{\rm P}$ being the reduced Planck
mass and a mass parameter  $m^2$ is introduced to make the action
dimensionless.  Greek indices run from 0 to 3 with conventions
$(-+++)$, while Latin indices run from 1 to 3, throughout the paper.
Further, we note that the Weyl-squared term ($\sqrt{-g}C^2$) is
invariant under the Weyl transformation of $g_{\mu\nu} \to
\tilde{g}_{\mu\nu}=\Omega^2(x) g_{\mu\nu}$. The Einstein equation is
given by
\begin{equation} \label{ein-eq}
G_{\mu\nu}+\kappa \Lambda g_{\mu\nu}-\frac{1}{m^2} B_{\mu\nu}= 0,
\end{equation} where the Einstein tensor $G_{\mu\nu}$ and  the Bach tensor $B_{\mu\nu}$ take the forms
\begin{eqnarray}
G_{\mu\nu}&=&R_{\mu\nu}-\frac{1}{2}Rg_{\mu\nu},\label{einstein} \\
 B_{\mu\nu}&=&2
\nabla^\rho\nabla^\sigma
C_{\mu\rho\nu\sigma}+G^{\rho\sigma}C_{\mu\rho\nu\sigma}.
\label{bach}
\end{eqnarray}
The solution is given by the de Sitter spacetime whose curvature
quantities are
\begin{equation}
\bar{R}_{\mu\nu\rho\sigma}=H^2(\bar{g}_{\mu\rho}\bar{g}_{\nu\sigma}-\bar{g}_{\mu\sigma}\bar{g}_{\nu\rho}),~~\bar{R}_{\mu\nu}=3H^2\bar{g}_{\mu\nu},~~\bar{R}=12H^2
\end{equation}
with $H^2=\kappa \Lambda/3$.  We describe  de Sitter spacetime
explicitly by choosing  either  a conformal time $\eta$ or a cosmic
time $t$
\begin{eqnarray} \label{frw1}
ds^2_{\rm dS}=\bar{g}_{\mu\nu}dx^\mu
dx^\nu&=&a(\eta)^2\Big[-d\eta^2+\delta_{ij}dx^idx^j\Big]\\
\label{frw2}&=&-dt^2+a^2(t)\delta_{ij}dx^idx^j,
\end{eqnarray}
where the conformal and  cosmic scale factors are given by
\begin{eqnarray}
a(\eta)=-\frac{1}{H\eta},~ a(t)=e^{Ht}.
\end{eqnarray}

The choice of Newtonian gauge $B=E=0 $ and $\bar{E}_i=0$  leads to
$10-4=6$ degrees of freedom (DOF)[Considering a fourth-order
equation for $h_{ij}$, it amounts  8 DOF]. In this case, the
cosmologically perturbed metric can be simplified to be
\begin{eqnarray} \label{so3-met}
ds^2=a(\eta)^2\Big[-(1+2\Psi)d\eta^2+2\Psi_i d\eta
dx^{i}+\Big\{(1+2\Phi)\delta_{ij}+h_{ij}\Big\}dx^idx^j\Big]
\end{eqnarray}
with the transverse vector $\partial_i\Psi^i=0$ and
transverse-traceless tensor $\partial_ih^{ij}=h=0$.
 In order to get
the cosmological perturbed equations,  one  is first to obtain the
bilinear action and then, varying it to yield the perturbed
equations.  We expand the Einstein-Weyl  gravity action (\ref{ECSW})
up to quadratic order in the perturbations of $\Psi,\Phi,\Psi_i,$
and $h_{ij}$ around  the de Sitter
background~\cite{Deruelle:2010kf}. Since the scalar, vector, and
tensor are decoupled from each other, we will consider the tensor
perturbed equation only
\begin{eqnarray}
&&\Box^2h_{ij}-m^2a^2\Box
h_{ij}+2m^2a^3Hh_{ij}^{\prime}=0\label{heq-2},
\end{eqnarray}
where $\Box=-d^2/d\eta^2+\partial_i\partial^i$ and the prime ($'$)
denotes the differentiation with respect to $\eta$. We introduce
Fourier modes for $h_{ij}$
\begin{eqnarray}\label{four-h}
h_{ij}(\eta,{\bf x})=\frac{1}{(2\pi)^{\frac{3}{2}}}\int d^3{k}\sum_{s={+,\times}}p_{ij}^{s}({\bf k})h_{\bf k}^{s}(\eta)e^{i{\bf
k}\cdot{\bf x}},
\end{eqnarray}
where  $p^{s}_{ij}$ is a linear polarization tensor with
$p^{+/\times}_{ij} p^{+/\times, ij}=1$ and  $h_{\bf k}^{+/\times}$
is a linearly polarized tensor mode. Plugging (\ref{four-h}) into
(\ref{heq-2}) leads to the fourth-order differential equation
\begin{eqnarray}
(h_{\bf k}^{s})^{''''}+2k^2(h_{\bf k}^{s})^{''}+k^4h_{\bf k}^{s}
+m^2a^2(h_{\bf k}^{s})^{''} +2m^2a^3H(h_{\bf
k}^{s})^{'}+m^2a^2k^2h_{\bf k}^{s}=0,\label{heq2}
\end{eqnarray}
which can be factorized to be
\begin{eqnarray}
\Big[\frac{d^2}{d\eta^2}-\frac{2}{\eta}\frac{d}{d\eta} +k^2\Big]
\Big[\eta^2\frac{d^2}{d\eta^2}-2\eta\frac{d}{d\eta}
+2+k^2\eta^2+\frac{m^2}{H^2} \Big]h_{\bf k}^{s}=0.\label{dec2}
\end{eqnarray}
In the limit of $m^2\to \infty$, one recovers the second-order
(tensor) equation for the Einstein gravity.  In the other limit of
$m^2\to 0$, one finds the fourth-order equation for the conformal
gravity
\begin{eqnarray}
\Big[\frac{d^2}{d\eta^2}-\frac{2}{\eta}\frac{d}{d\eta} +k^2\Big]
\Big[\eta^2\frac{d^2}{d\eta^2}-2\eta\frac{d}{d\eta} +2+k^2\eta^2
\Big]h_{\bf k}^{s}=0,\label{dec3}
\end{eqnarray}
which reduces further to a degenerate fourth-order
equation~\cite{Myung:2014cra}
\begin{eqnarray}
\Big[\frac{d^2}{d\eta^2} +k^2\Big]^2 h_{\bf k}^{s}=0
\xrightarrow[\eta\to z=-k\eta]{}\Big[\frac{d^2}{dz^2} +1\Big]^2
h_{\bf k}^{s}=0\label{dec4}
\end{eqnarray}
which is equivalent to the equation
\begin{equation}
\Box^2h_{ij}=0 \label{conf-h}
\end{equation}
for $h_{ij}$. Then, Eq.(\ref{dec3}) is considered as the Fourier
transform of the Weyl transformation for $\Box^2$:
\begin{equation}
\Box^2 \xrightarrow[\eta_{\mu\nu} \to
\bar{g}_{\mu\nu}=a^2\eta_{\mu\nu}]{}\Delta_4=\bar{\nabla}^2(\bar{\nabla}^2-2H^2),~~\bar{\nabla}^2=\frac{1}{\sqrt{-\bar{g}}}\partial_\mu(\sqrt{-\bar{g}}\bar{g}^{\mu\nu}\partial_\mu),
\end{equation}
where $\Delta_4$ is the fourth-order Lee-Wick operator in the dS
background. Also, we note the similar fourth-order equation (59) in
Ref.\cite{Mannheim:2011is} for the covariant approach to the
conformal gravity with the transverse gauge. Hence,
 the fourth-order equation (\ref{dec3}) can be interpreted
to be the tensor version of the Lee-Wick scalar equation.
Accordingly, Eq.(\ref{dec3}) implies two second-order equations with
$z=-k\eta=k/(aH)$
\begin{eqnarray}
\label{sec-dec1}\Big[\frac{d^2}{dz}-\frac{2}{z}\frac{d}{dz}
+1\Big]h_{\bf k}^{s,1}=0,\\
\Big[\frac{d^2}{dz^2}-\frac{2}{z}\frac{d}{dz}
+1+\frac{2}{z^2}\Big]h_{\bf k}^{s,2}=0\label{sec-dec2}
\end{eqnarray}
whose solutions to (\ref{sec-dec1}) and (\ref{sec-dec2}) are easily
found to be
\begin{eqnarray}
h_{\bf k}^{s,1}&=&c_1(i+z)e^{iz},\label{pmsol1}\\
h_{\bf k}^{s,2}&=&c_2ize^{iz},\label{pmmsol1}
\end{eqnarray}
where $c_{1}$ and $c_{2}$ are constants to be determined.
(\ref{pmsol1}) is the tensor solution to the Einstein gravity
[solution to a massless minimally coupled scalar], while
(\ref{pmmsol1}) is the other solution to the conformal gravity
[solution to a massless conformally coupled scalar] in de Sitter
spacetime. In this sense, we insist that the conformal gravity
includes the Einstein gravity as a part~\cite{Maldacena:2011mk}.

It is noted that (\ref{pmsol1}) and (\ref{pmmsol1}) are also the
solutions to the Lee-Wick scalar model~\cite{Myung:2014mla}. In the Lee-Wick scalar model, its
power spectrum  is Harrison-Zel'dovich scale-invariant as
\begin{equation} {\cal P}_{\rm LW}=\Big(\frac{H}{2\pi}\Big)^2,
\label{LW}
\end{equation}
whereas the spectrum for a massless minimally coupled scalar takes
the form
\begin{equation} {\cal
P}_{\rm
\phi}=\Big(\frac{H}{2\pi}\Big)^2\Big[1+\frac{k^2}{(aH)^2}\Big]=\Big(\frac{H}{2\pi}\Big)^2[1+z^2],
\label{mmcs}
\end{equation}
which is obviously a scale-variant spectrum and a scale-invariant superhorizon spectrum.

 The tensor power spectrum for the Einstein-Weyl gravity  was
given by~\cite{Deruelle:2012xv}
\begin{eqnarray} \label{ten-power1}
{\cal P}_{\rm EW}=\frac{ 2H^2}{\pi^2M^2_{\rm P}}\frac{m^2}{m^2+2H^2}
\Big(1+z^2-\frac{\pi}{2}z^3 |e^{i (\frac{\pi \nu}{2}+\frac{\pi}{4})}
H_{\nu}^{(1)}(z)|^2\Big),~~\nu=\sqrt{\frac{1}{4}-\frac{m^2}{H^2}}.
\end{eqnarray}
Here we worry about the negative-norm state because of minus $(-)$
sign in the front of the last term.

Fortunately, in the limit of $m^2\to0$ and $\kappa(=1/M^2_{\rm
P})\to\infty(m^2\kappa\to$ const), one finds the power spectrum
being  free from negative-norm state as
\begin{eqnarray} \label{ten-power1}
{\cal P}_{\rm EW}^{m^2\to0,\kappa\to\infty}=\frac{
m^2}{\pi^2M^2_{\rm P}}
\end{eqnarray}
which implies that the power spectrum  of conformal gravity is
properly recovered  from the Einstein-Weyl gravity when imposing
both $m^2\to0$ and $\kappa\to\infty$. In the next section, we will
obtain (\ref{ten-power1}) by computing the power spectrum of
conformal gravity itself.
\section{Conformal gravity}

In order to compute tensor power spectrum in the conformal
gravity~\cite{Myung:2014cra}, we begin with the action (\ref{CG}).
In this case, there is no restriction on $m^2$. Then, the
fourth-order differential equation is given by
\begin{eqnarray} \label{fourth-t}
\Box^2h_{ij}=0\to (h_{\bf k}^{s})^{''''}+2k^2(h_{\bf
k}^{s})^{''}+k^4h_{\bf k}^{s} =0,\label{heq2}
\end{eqnarray}
which is further rewritten  to be ({\ref{dec4}).
 This is  the same
equation for a degenerate Pais-Uhlenbeck (PU)
oscillator~\cite{Mannheim:2004qz,Kim:2013mfa} and its solution is
given by
\begin{equation} \label{desol}
h_{\bf k}^{s}(z)=\frac{N}{2k^2}\Big[i(a_2^s+a_1^s z)e^{iz}+c.c.\Big]
\end{equation}
 with $N$ the normalization constant. After quantization, $a^s_2$
 and $a^s_1$ are promoted to operators $\hat{a}^s_2({\bf k})$ and $\hat{a}^s_1({\bf
 k})$, which leads  to the expression $\hat{h}_{\bf k}^{s}(z)$.   The presence of $z$
in $(\cdots)$ reflects clearly that $ h_{\bf k}^{s}(z)$ is a
solution to the degenerate
 equation (\ref{dec4}). Together with $N=\sqrt{2\kappa m^2}$, the
canonical quantization could be
 accomplished by introducing
  commutation
relations between $\hat{a}_i^s({\bf k})$ and
$\hat{a}^{s\dagger}_j({\bf k}')$ as
 \begin{equation} \label{scft}
 [\hat{a}_i^s({\bf k}), \hat{a}^{s^{\prime}\dagger}_j({\bf k}')]= 2k \delta^{ss'}
 \left(
  \begin{array}{cc}
   0 & -i  \\
    i & 1 \\
  \end{array}
 \right)\delta^3({\bf k}-{\bf k}').
 \end{equation}
The tensor power spectrum is  defined by
\begin{eqnarray}\label{power}
\langle0|\hat{h}_{ij}(\eta,{\bf x})\hat{h}^{ij}(\eta,{\bf
x^{\prime}})|0\rangle=\int d^3{k}\frac{{\cal P}_{{\rm
h}}^{m^2}}{4\pi k^3}e^{i{\bf k}\cdot({\bf x}-{\bf x}')}.
\end{eqnarray}
Here we choose the Bunch-Davies vacuum $|0\rangle$ by imposing
$\hat{a}_i^s({\bf k})|0\rangle=0$. Using the definition
\begin{equation} \label{def-tps}
{\cal P}_{{\rm h}}^{m^2}e^{i{\bf k}\cdot({\bf x}-{\bf x}')}\equiv
\sum_{s,s'={+,\times}}{\cal P}^{ss',m^2}_{\rm h} \end{equation}
 and
substituting (\ref{four-h}) together with $\hat{h}_{\bf k}^{s}(z)$
into (\ref{power}), then one finds that ${\cal P}^{ss',m^2}_{\rm h}$
takes the form
\begin{eqnarray}
{\cal P}^{ss',m^2}_{\rm h}&=&\frac{m^2k}{4\pi^2M^2_{\rm P}}\int
d^3{\bf k}'\Big[\frac{1}{k'^2}p_{ij}^{s}({\bf k})p^{ijs'}({\bf k}')
\times\langle0|\Big([\hat{a}_2^s({\bf k}), \hat{a}^{s^{\prime}\dagger}_2({\bf k}')]+z[\hat{a}_2^s({\bf k}), \hat{a}^{s^{\prime}\dagger}_1({\bf k}')]\nonumber\\
&&\hspace*{7em}+z[\hat{a}_1^s({\bf k}),
\hat{a}^{s^{\prime}\dagger}_2({\bf k}')]+z^2[\hat{a}_1^s({\bf k}),
\hat{a}^{s^{\prime}\dagger}_1({\bf k}')]\Big)
|0\rangle e^{i({\bf k}\cdot{\bf x}-{\bf k'}\cdot{\bf x}^{\prime})}\Big]\label{spss}\\
&=&\frac{m^2k}{4\pi^2M^2_{\rm P}}\int d^3{\bf
k}'\Big[\frac{1}{k'^2}p_{ij}^{s}({\bf k})p^{ijs'}({\bf k}')
\times\langle0|[\hat{a}_2^s({\bf k}),
\hat{a}^{s^{\prime}\dagger}_2({\bf k}')]|0\rangle e^{i({\bf
k}\cdot{\bf x}-{\bf k'}\cdot{\bf x}^{\prime})}\Big]\\
&=&\frac{m^2}{2\pi^2M^2_{\rm
P}}p_{ij}^{s}p^{ijs'}\delta^{ss'}e^{i{\bf k}\cdot({\bf x}-{\bf
x}')}\label{pss}.
\end{eqnarray}
In obtaining (\ref{pss}), we used the commutation relations of
(\ref{scft}) which reflect the quantum nature of a degenerate PU
oscillator. This is in contrast to a non-degenerate PU oscillator
for the Lee-Wick scalar theory~\cite{Myung:2014mla}.

Finally, from (\ref{def-tps}) and (\ref{pss}),  we obtain the tensor
power spectrum
\begin{equation} \label{tensorp11}
{\cal P}^{ m^2}_{\rm h}=\frac{m^2}{\pi^2M^2_{\rm P}},
\end{equation}
which is the same form as in (\ref{ten-power1}).

 Further, by
choosing $m^2=2H^2$ specifically, (\ref{tensorp11}) reduces to
\begin{equation} \label{sca-inv-t}
{\cal P}^{m^2=2H^2}_{\rm h}=\frac{2H^2}{\pi^2M^2_{\rm P}},
\end{equation}
which is  the scale-invariant tensor spectrum. This is surely
compared to the scale-variant tensor spectrum for the Einstein
gravity~\cite{Baumann:2009ds}
\begin{equation} \label{ein-ts}
P_{\rm h}(k)=\frac{2H^2}{\pi^2M^2_{\rm
P}}\Big(1+\frac{k^2}{(aH)^2}\Big)
\end{equation}
during de Sitter inflation. It is worth noting  that one could not
find the scale-invariant form of (\ref{sca-inv-t})  from the general
expression (\ref{tensorp11}) unless one chooses $m^2=2H^2$. This
implies  that there is an ambiguity in determining  a correct
amplitude of the power spectrum for conformal gravity.
\section{Discussions}

We have found the scale-invariant tensor power spectrum generated
during de Sitter inflation from conformal gravity with mass
parameter $m^2=2H^2$. This scale-invariant tensor spectrum  could be
understood because the conformal gravity is invariant under the Weyl
transformation. This contrasts to the scale-variant  tensor spectrum
(\ref{ein-ts})  of Einstein gravity which is not Weyl-invariant.
This is considered as a tensor version of Harrison-Zel'dovich
scale-invariant spectrum obtained from the Lee-Wick scalar model
with mass parameter $M^2=2H^2$. We summarize below the difference
between second-order theory and fourth-order theory and similarity
between scalar and tensor in the same order. Their DOF are shown
explicitly.
\begin{center}
\begin{tabular}{|c|c|c|}
  \hline power spectrum & scalar theory (DOF)  & tensor theory (DOF) \\
  \hline
  scale-variant & massless minimally coupled  & Einstein gravity (2)\\
 (second-order) & scalar theory (1) &  \\ \hline
  scale-invariant & Lee-Wick model & Conformal gravity \\
 (fourth-order)  &  with $M^2=2H^2$ (2) &  with $m^2=2H^2$ (4)\\
  \hline
\end{tabular}
\end{center}

\vspace{0.5cm} Finally, we would like to explain  why our
computation is meaningful even in the presence of ghost. As was
shown in (\ref{fourth-t}), the conformal gravity provides a healthy
massless spin-2 and an unhealthy (ghost) massless spin-2.  In
computing the power spectrum, one has to distinguish   higher-time
derivative terms (ghost) from ghost state (negative-norm state). In
the case of the Lee-Wick model, its scalar power spectrum based on
the quantization scheme of  non-degenerate PU oscillator is given by
${\cal P}_{\rm LW}=(H/2\pi)^2[1+(k\eta)^2-(k\eta)^2]$ where the
first two are contribution from a massless minimally scalar and the
last from a massless conformally coupled scalar when fourth-order LW
operator is partially factorized into two second-order operators
with minus sign in  front of the  massless conformally coupled
scalar operator. Here there is no negative-norm (ghost state) for
${\cal P}_{\rm LW}$ because the second term cancels against the
third term. This explains why the LW computation is meaningful even
in the presence of ghost. Similarly, the power spectrum of conformal
gravity based on the quantization scheme (\ref{scft}) of degenerate
PU oscillator (\ref{desol})  is given by (\ref{tensorp11}) which is
free from negative-norm state (ghost state) and scale-invariant.
This implies that off-diagonal cancellation of  $z$-order in
(\ref{spss}) using (\ref{scft}) is  important to have positive-norm
state.  Consequently, our whole computation is safe even in the
presence of ghost because the power spectrum of conformal gravity
(\ref{tensorp11}) [equivalently, (\ref{ten-power1})]  is free from
ghost state.

\vspace{0.25cm}

 {\bf Acknowledgement}

\vspace{0.25cm}
 This work was supported by the National
Research Foundation of Korea (NRF) grant funded by the Korea
government (MEST) (No.2012-R1A1A2A10040499).

\end{document}